\documentclass[]{spie}  

\usepackage{booktabs}
\usepackage{multirow}
\usepackage{relsize}
\usepackage{amsmath,amsfonts,amssymb}
\usepackage{graphicx}
\usepackage[colorlinks=true, allcolors=blue]{hyperref}
\newcommand{\ul}[1]{\underline{#1}}

\newcommand\blfootnote[1]{%
  \begingroup
  \renewcommand\thefootnote{}\footnote{#1}%
  \addtocounter{footnote}{-1}%
  \endgroup
}

\title{Efficient motion-based metrics for video frame interpolation}

\author{Conall Daly}
\author{Darren Ramsook}
\author{Anil Kokaram}
\affil{Sigmedia Group,  Electronic and Electrical Engineering Dept., Trinity College Dublin, \\ Dublin, Ireland}

\begin{document} 
\maketitle
\blfootnote{For further information contact Conall Daly: \href{mailto:dalyc21@tcd.ie}{dalyc21@tcd.ie}}
\begin{abstract}
Video frame interpolation (VFI) offers a way to generate intermediate frames between consecutive frames of a video sequence. Although the development of advanced frame interpolation algorithms has received increased attention in recent years, assessing the perceptual quality of interpolated content remains an ongoing area of research. In this paper, we investigate simple ways to process motion fields, with the purposes of using them as video quality metric for evaluating frame interpolation algorithms. We evaluate these quality metrics using the BVI-VFI dataset which contains perceptual scores measured for interpolated sequences. From our investigation we propose a motion metric based on measuring the divergence of motion fields. This metric correlates reasonably with these perceptual scores (PLCC=0.51) and is more computationally efficient ($\times2.7$ speedup) compared to FloLPIPS (a well known motion-based metric). We then use our new proposed metrics to evaluate a range of state of the art frame interpolation metrics and find our metrics tend to favour more perceptual pleasing interpolated frames that may not score highly in terms of PSNR or SSIM.
\end{abstract}


\section{Introduction}
\label{sec:intro}

Video frame interpolation (VFI) offers a way to generate intermediate frames between consecutive frames of a video sequence. By increasing the effective frame rate, VFI can enhance the viewing experience and improve downstream tasks such as slow-motion playback, virtual reality, and frame rate up-conversion for broadcast. Numerous VFI methods have emerged in recent years, powered by advanced optical flow algorithms and deep neural networks that can effectively estimate complex, non-rigid motions \cite{jin2023unified,park2021abme,niklaus2020softmax,danier2023ldmvfi}.

Despite the remarkable progress in interpolation quality, the evaluation and comparison of VFI methods remain non-trivial. Traditional full-reference metrics such as Peak Signal-to-Noise Ratio (PSNR) and Structural Similarity (SSIM) are widely used but are often criticized for their limited ability to capture temporal consistency, an essential aspect of perceived video quality. Meanwhile, more perceptually aligned metrics like LPIPS\cite{zhang2018unreasonable} have been adapted to the video domain, such as FloLPIPS \cite{danier2022flolpips}, to better reflect how human observers perceive temporal artifacts. Temporal consistency is also an important factor in the quality of videos created by deep generative video models and is still an open problem \cite{unterthiner2019fvd,songwei2024fvdbias,liu2024frechet}.

However, these metrics can be computationally expensive. The authors of FloLPIPS noted that while it is a differentiable quality metric it is quite computationally expensive. This makes it unsuitable for training a neural network for the purposes of frame interpolation or other video generation task. In addition, further work using a subjective study of frame interpolation algorithms performed an extensive test of how different image and video based metrics correlate to perceptual scores. They found that most video quality metrics correlated rather poorly, including FloLPIPS. The metric with the highest overall correlation was a classical motion based metric, full-reference assessor along salient trajectories (FAST) \cite{wu2019quality}. FAST tracks feature points (e.g. Scale Invariant Features) through a subset of frames in a reference and distorted (interpolated) sequence. Small patches around these features are extracted from both the motion field and frames themselves. The authors build a histogram for each of the motion fields and measure the distance between them. This is combined this with an image quality measure, gradient magnitude similarity deviation (GMSD) \cite{xue2013gradient}. The high correlation of this metric shows that by thinking carefully about processing of motion field features video quality metrics can be developed that beat even deep perceptual metrics. Our work aims to investigate novel ways of processing motion fields for this purpose.

In this paper, we explore whether motion fields—obtained from both the original high frame rate (HFR) video (or ground-truth references) and the interpolated (distorted) sequence—can be used as a lightweight, effective quality metric for evaluating VFI. Leveraging standard optical flow estimators and concepts such as End Point Error (EPE), Temporal Smoothness (TS), and Vector-Median End Point Error (VM-EPE), we propose to measure how closely the motion in the interpolated frames agrees with a reference motion field. By doing so, we aim to capture not only spatial fidelity but also the temporal coherence necessary for high-quality video interpolation. We compare our motion-based metrics with popular baselines such as PSNR, SSIM, and FloLPIPS, and we examine their correlations with subjective assessments. Ultimately, our goal is to determine if motion-field metrics can provide a computationally efficient, interpretable, and sufficiently accurate measure of VFI performance. In summary, our main contributions are
\begin{itemize}
    \item Proposing a number of novel motion-based metrics 
    \item Showing that across multiple resolutions and frame-rates quality metric performance can vary drastically, so there is no one size fits all approach
    \item A novel motion weighted image metric that has low computational complexity while remaining reasonably correlated to perceptual scores at a resolution of 1080p
\end{itemize}

\section{Background}
\label{sec:background}
This section examines existing approaches to quality assessment in VFI and explores how temporal consistency metrics can serve as effective video quality measures for VFI applications.

\subsection{Quality Assessment Metrics for Video Frame Interpolation}
\label{sec:vfi-qual-metrics}
Video quality assessment methods are classified into three categories: no-reference (NR), reduced-reference (RR), and full-reference (FR). The effectiveness of these metrics is commonly assessed through their correlation with human perceptual judgments, utilizing statistical measures including Pearson's Linear Correlation Coefficient (PLCC), Spearman's Rank Correlation Coefficient (SRCC), Kendall's Rank Correlation Coefficient (KRCC), and Root Mean Squared Error (RMSE), following VQEG standardization \cite{vqeg2000final}.

Contemporary video quality metrics increasingly incorporate spatio-temporal characteristics \cite{han2025full,ccokmez2025clip}. ST-GREED \cite{madhusudana2021stgreed} examines spatial and temporal band-pass coefficients through Generalized Gaussian Distributions (GGD), quantifying entropy variations between reference and interpolated videos to identify distortions arising from frame rate modifications and compression artifacts. Alternatively, Wu \textit{et al.} \cite{wu2019quality} utilize estimated motion directly by tracking key-point trajectories and extracting motion field patches in their vicinity. They assess temporal consistency deviations by comparing histograms of these patches between reference and interpolated sequences.

Nevertheless, conventional video metrics frequently show poor correlation with subjective assessments for VFI-specific artifacts \cite{men2020visual}, leading to the development of specialized approaches. Hou \textit{et al.} \cite{hou2022perceptual} implement a Swin Transformer-based framework trained explicitly on subjective ratings from their proprietary dataset (VFIPS). This method demonstrates strong correlation performance with PLCC of 0.87 and SRCC of 0.79. However, its computational demands are substantial, requiring 2.5 GB of GPU memory for processing a 256×256 12-frame sequence.

Danier \textit{et al.} introduce FloLPIPS, a video quality metric that extends LPIPS through the integration of temporal distortion sensitivity. LPIPS quantifies perceptual differences between reference ($\mathbf{I}$) and interpolated ($\hat{\mathbf{I}}$) frames using deep feature representations $\phi_l(\hat{\mathbf{I}})$, $\phi_l(\mathbf{I})$ extracted from layer $l$ of a pre-trained convolutional neural network. At each spatial location $\mathbf{x}$, LPIPS computes $\|w_l \odot (\phi_l(\hat{\mathbf{I}})-\phi_l(\mathbf{I}))\|_2^2$ across channels, where $w_l$ represents a weighting factor that emphasizes or de-emphasizes features from layer $l$. These differences are then spatially averaged. FloLPIPS extends LPIPS by incorporating motion error weighting. For consecutive frame pairs $\hat{\mathbf{I}}_{n-1}$, $\hat{\mathbf{I}}_n$ and $\mathbf{I}_{n-1}$, $\mathbf{I}_n$, it estimates motion fields $\hat{\mathbf{F}}$ and $\mathbf{F}$ using a pre-trained optical flow estimator. The endpoint error (EPE) between these motion fields is then computed.

\begin{equation}
\Delta \mathbf{F}(\mathbf{x}) = \|\hat{\mathbf{F}}(\mathbf{x}) - \mathbf{F}(\mathbf{x})\|_2
\end{equation} 

This EPE map undergoes spatial normalization to generate distortion-sensitive weights.

\begin{equation}
\label{eq:flow-weight}
\mathbf{w}(\mathbf{x}) = \frac{\Delta \mathbf{F}(\mathbf{x})}{\sum_{\mathbf{x}} \Delta \mathbf{F}(\mathbf{x})}
\end{equation}

These weights modify LPIPS's uniform spatial averaging approach, placing greater emphasis on regions exhibiting motion inconsistencies. The final FloLPIPS score combines the weighted feature differences across all layers and frames.

\begin{equation}
\text{FloLPIPS} = \sum_l \sum_{\mathbf{x}} \mathbf{w}(\mathbf{x}) \|w_l \odot (\phi_l(\hat{\mathbf{I}})-\phi_l(\mathbf{I}))\|_2^2
\end{equation}

FloLPIPS demonstrates performance scores of 0.71 PLCC and 0.68 SRCC on the BVI-VFI dataset. However, FloLPIPS presents significant computational overhead. The reported runtime is 332 ms for processing a 1080p frame compared to LPIPS's 59 ms, representing a 5.5$\times$ performance penalty. The effectiveness of FloLPIPS stems from its use of EPE as a temporal consistency measure. Consequently, we propose adopting a simpler, more computationally efficient approach to temporal consistency assessment.

\section{Proposed Metrics}
\label{sec:prop-metrics}

Considering a current frame, $I_t^{ref}$, from a high frame rate (HFR) video, and a future frame $I_{t+1}^{ref}$ as our reference sequence. We have the same frames generated by a frame interpolation algorithm, $I_t^{dis}$ and $I_{t+1}^{dis}$, of course only half the frames will be interpolated and inserted into the low frame rate (LFR) sequence. We compute the motion field for each of these sequences $F_{t\rightarrow t+1}^{ref}$ and $F_{t\rightarrow t+1}^{dis}$.

\begin{equation}
\centering
    F_{t\rightarrow t+1}^{ref} = \text{MotionEstimator}(I_t^{ref}, I_{t+1}^{ref})
\end{equation}

\begin{equation}
\centering
    F_{t\rightarrow t+1}^{dis} = \text{MotionEstimator}(I_t^{dis}, I_{t+1}^{dis})
\end{equation}

\subsection{Measures of Temporal Consistency}

A key aspect of evaluating frame interpolation quality is assessing the temporal consistency of motion across frames. We achieve this by comparing the motion fields of a high frame rate (HFR) reference sequence with those generated for the interpolated sequence. We define the motion fields for the reference and interpolated sequences as $F_{t\rightarrow t+1}^{ref}=(u^{ref},v^{ref})$ and $F_{t\rightarrow t+1}^{dis}=(u^{dis},v^{dis})$, respectively, where $u$ and $v$ represent the horizontal and vertical motion components.

\subsubsection{End Point Error}

Our first metric is the popular End Point Error (EPE), which measures the Euclidean distance between corresponding motion vectors in the reference and interpolated motion fields.

\begin{equation}
\label{eq:epe}
    \text{EPE} = \frac{1}{N\times M}\mathlarger{\sum}_{j=1}^{N\times M} \sqrt{\left(u^{ref}_j-u^{dis}_j\right)^2 + \left(v^{ref}_j-v^{dis}_j\right)^2}
\end{equation}

We compute the mean EPE over a frame of height $N$ and width $M$, then average across frames to obtain an EPE measure for the entire sequence. Note we assume a dense motion field with one vector per pixel. This serves as a meaningful metric for frame interpolation, as it quantifies discrepancies in motion fields. A lower EPE indicates that the interpolated motion is closely aligned with the ground truth, preserving temporal consistency.

\subsubsection{Temporal Smoothness}
In addition to measuring direct motion differences, we evaluate motion-compensated motion differences to assess the temporal smoothness of motion trajectories across interpolated frames. This approach treats motion vectors as trajectories, where a motion vector at time $t$, $d^t(x,y)=(u^t,v^t)$, is used to locate the corresponding position in the next motion field at $t+1$: $d^{t+1}(x+u^t,y+v^t)=(u^{t+1},v^{t+1})$. We define our Temporal Smoothness (TS) metric as:

\begin{equation}
    \text{TS} = \frac{1}{N\times M}\mathlarger{\sum}_{j=1}^{N\times M} \sqrt{(u^t_j -u^{t+1}_j)^2 + (v^t_j - v^{t+1}_j)^2}
\end{equation}

This is a no-reference metric which evaluates how consistently motion evolves over time, ensuring that interpolated motion remains temporally coherent. A lower TS value indicates smoother transitions between motion fields.

\subsection{Measures of Spatial Consistency}
While temporal consistency ensures motion coherence across frames, spatial consistency evaluates the smoothness and regularity of motion fields within individual frames. Frame interpolation algorithms can introduce spatial artifacts such as motion discontinuities, flow singularities, or irregular motion patterns that violate the natural spatial structure of optical flow. These artifacts manifest as abrupt changes in motion vectors between neighbouring pixels. To detect and quantify such spatial inconsistencies, we propose several metrics that assess motion field smoothness, detect flow singularities, and evaluate deviations from expected spatial motion patterns

\subsubsection{Vector Median Motion Field Filtering}
To evaluate the spatial consistency of motion fields, we compute vector median filtered versions of both the reference and distorted motion fields, denoted as $F_{t\rightarrow t+1}^{ref,VM}$ and $F_{t\rightarrow t+1}^{dis,VM}$. The vector median is computed over an $n\times n$ patch of the motion field. For each motion vector $d_i=(u_i,v_i)$ within the patch, we compute its distance energy, which is the sum of Euclidean distances to all other motion vectors $d_j=(u_j,v_j)$ in the patch:

\begin{equation}
    E(d_i) = \mathlarger{\sum}_{j=1}^{n\times n} \sqrt{(u_i-u_j)^2 + (v_i-v_j)^2}
\end{equation}

The vector median, which provides a robust estimate of the motion for that patch, is the vector $d_i$ that minimizes this distance energy:

\begin{equation}
    d^{VM} = \underset{d_i}{argmin} \; E(d_i)
\end{equation}

This process yields a new flow field with a smoothness prior enforced on the original flow field. We use this smoothed motion field as a robust estimate of the true underlying motion for a reference frame.

\subsubsection{Vector Median End Point Error}
We compute the EPE between the interpolated sequence motion field, $F_{t\rightarrow t+1}^{dis}$, and its vector median filtered version, $F_{t\rightarrow t+1}^{dis,VM}$, calling this metric VM-EPE. This measures how much the original motion field deviates from its spatially smoothed version.

\subsubsection{Motion Field Smoothness Dissimilarity}
We also compare the smoothness characteristics between reference and distorted motion fields by computing the difference between their respective VM-EPE values:

\begin{equation}
    S_{dis}-S_{ref} = || \text{VM-EPE}^{dis} - \text{VM-EPE}^{ref} ||
\end{equation}

This metric captures how differently the reference and distorted sequences behave in terms of spatial motion consistency.

\subsubsection{Motion Field Divergence}
We take inspiration from the literature on the restoration of archival footage. Corrigan et al. developed techniques for the handling of pathological motion \cite{corrigan2006pathological} in blotch detection. We measure the absolute divergence of the flow field for the distorted sequence to detect flow singularities. 

\begin{equation}
    \left| \nabla \cdot F_{t\rightarrow t+1}^{dis} \right| = \left| \frac{\partial F_{t\rightarrow t+1}^{dis,x}}{\partial x} + \frac{\partial F_{t\rightarrow t+1}^{dis,y}}{\partial y} \right|
\end{equation}

We refer to this no-reference metric as DIV. A value of $\left| \nabla \cdot F_{t\rightarrow t+1}^{dis} \right|$ close to $0$ indicates that the motion field is not flowing towards or away from one point. This provides a useful measure of motion field smoothness.

\subsection{Augmenting Image-Based Quality Measures}
Previous works have used motion to weight image metrics in order to create a combined spatio-temporal metric for a video sequence. Often, this is achieved simply by multiplying a motion-based metric by an image metric. We use a selection of our motion metrics to weight both PSNR and SSIM which augments these image-based quality metrics with motion quality information. We weight the PSNR or SSIM measured over the whole frame. Intuitively, we wish to fallback to the image metric when the motion integrity over the whole frame is good. Conversely we penalise the metric when that motion integrity is poor, because the underlying image would probably be of poor quality. Using $\alpha$ to represent the motion error (EPE, TS or DIV discussed above) we can define a weight $w$ as follows, in the case of EPE
\begin{equation}
    w_{EPE} = \frac{1}{1 + \alpha_{EPE}}
\end{equation}
This means that for large motion errors the weighting is close to 0 and for small motion errors the weighting is close to 1. It also maps our motion metric from $[0,\infty)$ to $[0,1)$. In what follows we denote a motion-weighted metric using subscript EPE, TS or DIV

\begin{equation}
    PSNR_{EPE} = w_{EPE} \cdot PSNR
\end{equation}

\section{Experimental Setup} 
This section outlines our experimental methodology for evaluating the proposed motion-based quality metrics and their application to VFI algorithm assessment. We detail the dataset selection, evaluation protocols, and the range of quality metrics and interpolation algorithms used for comprehensive comparative analysis.

\subsection{Evaluation Dataset} 
We conduct our evaluation using the BVI-VFI database, which represents the sole publicly accessible subjective quality dataset featuring uncompressed video sequences in 8-bit YUV420 format, with distortions generated exclusively through video frame interpolation (VFI) processes. When compared to the two other predominantly utilized VFI quality assessment datasets, VFIPS and VFIIQA, the BVI-VFI database uniquely provides Differential Mean Opinion Scores (DMOS), establishing it as the most appropriate choice for full-reference (FR) metric evaluation and validation. The comprehensive BVI-VFI dataset encompasses three distinct spatial resolutions (540p, 1080p, 2160p) alongside three different frame rates (30, 60, 120fps), thereby providing varied content for investigating perceptual artifacts across multiple viewing conditions. For the purposes of this investigation, we concentrate on a carefully selected subset that includes 540p and 1080p resolutions operating at 30 and 60fps, as these configurations represent more prevalent resolution and framerate conversion scenarios encountered in practical applications, while simultaneously presenting greater challenges for existing quality assessment metrics.

\subsection{Evaluation Metrics}
Following the standard methodology recommended by ITU-T Rec. P.1401, we measure the performance of quality metrics using Pearson's Linear Correlation Coefficient (PLCC) to assess linearity, Root Mean Squared Error (RMSE) to evaluate prediction accuracy, and both Kendall's Rank Correlation Coefficient (KRCC) and Spearman's Rank Correlation Coefficient (SRCC) to determine the monotonicity of predictions. For the computation of these statistics, we first fit a logistic function between the calculated quality indices and the DMOS values according to the model specified by the Video Quality Expert Group\cite{vqeg2000final}. This model accounts for the non-linear relationship between objective measurements $x$ and subjective perception estimated as $Y(x)$ as follows.

\begin{equation}
Y(x) = \beta_2 + \frac{\beta_1-\beta_2}{1+\exp\left(-\frac{x-\beta_3}{|\beta_4|}\right)}
\end{equation}

where $\beta_k$ are model parameters estimated using BFGS fitting\cite{}.

\subsection{Compared Quality Metrics}
We compare our metrics against the two most widely used image metrics PSNR and SSIM. We also compare against the deep perceptual metric FlolPIPS. We use FloLPIPS with its default components PWC-Net \cite{sun2018pwcnet} (motion estimator) and  AlexNet \cite{krizhevsky2012alex} (feature extractor).

\subsection{Compared Frame Interpolation Algorithms} The algorithms evaluated in Section \ref{subsec:eval-vfi} are the five compared in BVI-VFI\cite{} along with a selection of classical and deep learning based frame interpolators. We use the ffmpeg implementation (Minterpolate) of Choi et al.'s \cite{choi2007motion} and Kokaram et al's Bayesian based frame interpolator ACKMRF \cite{kokaram2020bayesian},  as examples of widely used classical frame-interpolators. We use the following deep learning based frame interpolators, ABME \cite{park2021abme}, revisiting-sepconv \cite{niklaus2021revisiting}, softmax-splatting \cite{niklaus2020softmax}, UPR-Net \cite{jin2023unified}, RIFE\cite{}, FILM \cite{reda2022film}, the diffusion based LDMVFI \cite{danier2023ldmvfi}.

\section{Results and Discussion}
\label{sec:results}

\subsection{Correlation to Perceptual Data}

\begin{table}[]
\centering

\caption{\textit{The performance of the tested quality metrics on the DMOS values in BVI-VFI dataset across different frame rates and interpolation methods. In each column, the metrics most highly correlated with DMOS are \textbf{boldfaced} and the second most highly correlated are \underline{underlined}.}}
\label{tab:corr_tab}

\hfill

\resizebox{\columnwidth}{!}{
\begin{tabular}{@{}l||ll|ll||ll|ll||llll@{}}
\toprule
         & \multicolumn{4}{c||}{540p}                                     & \multicolumn{4}{c||}{1080p}                                    & \multicolumn{4}{c}{\multirow{2}{*}{Overall}}                   \\ \cmidrule(lr){2-9}
         & \multicolumn{2}{c|}{15fps $\rightarrow$ 30fps}     & \multicolumn{2}{c||}{30fps $\rightarrow$ 60fps}    & \multicolumn{2}{c|}{15fps $\rightarrow$ 30fps}     & \multicolumn{2}{c||}{30fps $\rightarrow$ 60fps}    & \multicolumn{4}{c}{}                                           \\ \midrule
Metric   & PLCC          & SRCC         & PLCC          & SRCC         & PLCC          & SRCC         & PLCC          & SRCC         & PLCC          & KRCC          & SRCC         & RMSE           \\ \midrule
PSNR                        & 0.41                        & 0.27                       & 0.35                  & 0.45                              & 0.40                        & 0.45                       & 0.61                  & 0.66                              & 0.42          & 0.32          & 0.46          & 16.70          \\
SSIM                        & 0.40                        & 0.35                       & 0.48                  & \textbf{0.60}                     & 0.42                        & 0.44                       & 0.62                  & {\ul{0.71}}                        & {\ul{0.48}}    & {\ul{0.36}}    & {\ul{0.53}}    & {\ul{16.06}}    \\
FloLPIPS                    & \textbf{0.49}               & {\ul{0.42}}                 & {\ul{0.53}}            & {\ul{0.55}}                        & \textbf{0.60}               & \textbf{0.55}              & \textbf{0.78}         & \textbf{0.74}                     & \textbf{0.58} & \textbf{0.40} & \textbf{0.58} & \textbf{14.98} \\ \midrule
EPE                         & {\ul{0.48}}                  & \textbf{0.44}              & 0.51                  & \textbf{0.60}                     & 0.35                        & 0.37                       & 0.54                  & 0.61                              & 0.46          & 0.34          & 0.45          & 16.32          \\
TS                      & 0.45                        & 0.40                       & \textbf{0.54}         & 0.54                              & 0.31                        & 0.31                       & 0.56                  & 0.63                              & 0.44          & 0.37          & 0.46          & 16.51          \\
DIV                         & 0.32                        & 0.29                       & 0.50                  & 0.39                              & {\ul{0.51}}                  & {\ul{0.53}}                 & {\ul{0.68}}            & 0.47                              & 0.47          & 0.27          & 0.40          & 16.19          \\
VM-EPE                      & 0.30                        & 0.26                       & 0.48                  & 0.38                              & 0.34                        & 0.30                       & {\ul{0.68}}            & 0.57                              & 0.45          & 0.25          & 0.38          & 16.40          \\
S\textsubscript{dis}-S\textsubscript{ref}               & 0.31                        & 0.29                       & 0.52                  & 0.46                              & 0.33                        & 0.30                       & 0.57                  & 0.56                              & 0.45          & 0.29          & 0.43          & 16.42           \\ \bottomrule
\end{tabular}
}
\end{table}

Table \ref{tab:corr_tab} presents the correlation performance of our proposed motion-based metrics alongside established image quality metrics (PSNR and SSIM) and the state-of-the-art FloLPIPS on the BVI-VFI dataset. The results reveal several key insights regarding the effectiveness of different metric categories across varying resolution and frame rate conditions.

FloLPIPS consistently demonstrates superior performance across most test conditions, achieving the highest overall correlation scores (PLCC=0.58, SRCC=0.58) and the lowest RMSE (14.98). This validates the effectiveness of incorporating motion information into perceptual quality assessment for VFI applications. However, the performance gap between FloLPIPS and our simpler motion-based metrics varies significantly across different test scenarios, suggesting that computational complexity may not always translate to proportional quality assessment improvements.

The results exhibit a clear resolution dependency, with all metrics showing substantially higher correlation scores for 1080p sequences compared to 540p sequences. This trend is particularly pronounced for FloLPIPS, which achieves correlation scores of 0.60-0.78 (PLCC) for 1080p sequences versus 0.49-0.53 for 540p sequences. This performance degradation at lower resolutions likely stems from FloLPIPS's reliance on pre-trained networks (PWC-Net for motion estimation and AlexNet for feature extraction) that may not have been optimally trained for lower resolution content.

The TS metric shows particularly strong performance for 540p sequences at 30$\rightarrow$60fps conversion (PLCC=0.54), where it achieves the highest correlation among all tested metrics. This indicates that evaluating motion trajectory consistency becomes especially important when interpolating at higher frame rates, where temporal artifacts are more perceptually significant.

\begin{figure}
    \centering
    \begin{tabular}{cc}
       \includegraphics[width=0.4\linewidth]{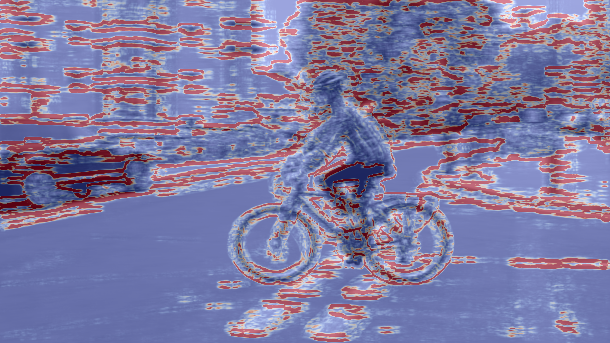}  & 
       \includegraphics[width=0.4\linewidth]{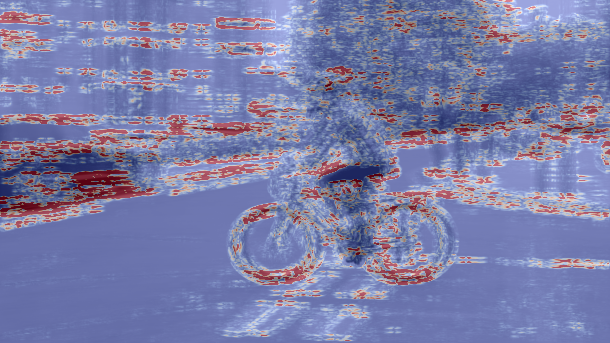} \\
       Ground Truth & Average \\\\
       \includegraphics[width=0.4\linewidth]{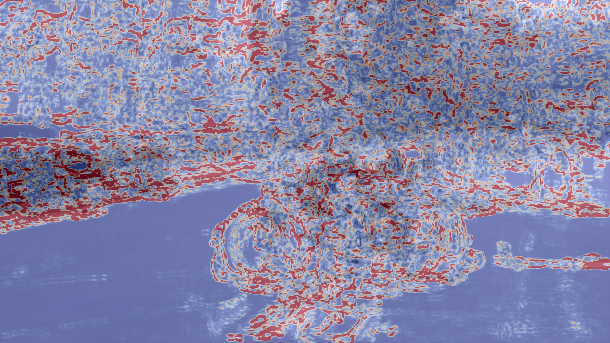} &
       \includegraphics[width=0.4\linewidth]{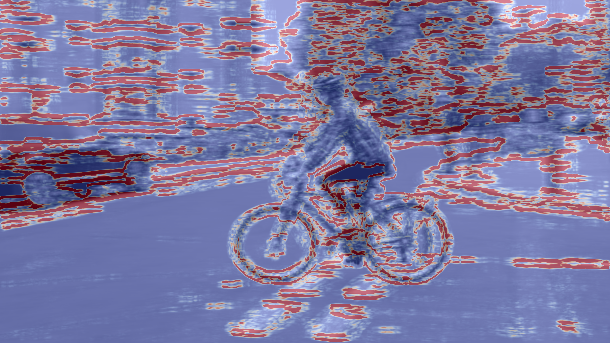} \\
       DVF & ST-MFNet \\\\
    \end{tabular}
    \caption{\textit{Visualisation of DIV metric for a frame from the BVI-VFI dataset. Values close to 0 (smooth motion field) are shown in {\color{blue}blue} and values close to 1 (distorted motion field) are shown in {\color{red}red}. It can be seen the DIV is effective in highlighting areas where VFI algorithms may struggle, such as the occlusion and uncovering around the bike wheel. It is also able to highlight regions of motion inconsistency as seen in the heavy warping of DVF.}}
    \label{fig:div-vis}
\end{figure}

The DIV metric demonstrates notable performance for higher resolution sequences, achieving second-best correlation scores for 1080p content (PLCC=0.51-0.68). This suggests that motion field divergence serves as an effective indicator of spatial motion consistency. A plot of DIV overlayed on a frame from a sequence within BVI-VFI is shown in Figure \ref{fig:div-vis}. As can be seen DIV seems to be able to highlight areas of motion inconsistency within a frame which could help in weighting image errors.

The relatively poor performance of VM-EPE and S\textsubscript{dis}-S\textsubscript{ref} indicates that vector median filtering may introduce excessive smoothing that removes perceptually relevant motion details.

The results reveal interesting patterns across different frame rate conversion scenarios. For 15$\rightarrow$30fps conversion, EPE consistently performs well across both resolutions, suggesting that this metric is particularly effective for detecting artifacts in moderate frame rate upsampling. Conversely, for 30$\rightarrow$60fps conversion, the performance differences between metrics become more pronounced, with FloLPIPS showing significantly better correlation for 1080p sequences while our proposed metrics maintain competitive performance for 540p content.

PSNR and SSIM, while computationally efficient, consistently underperform compared to motion-aware metrics across all test conditions. SSIM shows better correlation than PSNR, particularly for higher frame rate conversions (30$\rightarrow$60fps), but both metrics fail to capture the temporal artifacts that are characteristic of VFI applications. The competitive performance of simpler motion-based metrics like EPE suggests that motion is the dominant component in effective VFI quality assessment.

\begin{table}[]
\centering

\caption{\textit{The performance of the motion-weighted image quality metrics on the DMOS values in BVI-VFI dataset across different frame rates and interpolation methods. In each column, the metrics most highly correlated with DMOS are \textbf{boldfaced} and the second best are \underline{underlined}.}}
\label{tab:augmented-image}

\hfill

\resizebox{\columnwidth}{!}{
\begin{tabular}{@{}c||cc|cc||cc|cc||cccc@{}}
\toprule
         & \multicolumn{4}{c||}{540p}                                     & \multicolumn{4}{c||}{1080p}                                    & \multicolumn{4}{c}{\multirow{2}{*}{Overall}}                   \\ \cmidrule(lr){2-9}
         & \multicolumn{2}{c|}{15fps $\rightarrow$ 30fps}     & \multicolumn{2}{c||}{30fps $\rightarrow$ 60fps}    & \multicolumn{2}{c|}{15fps $\rightarrow$ 30fps}     & \multicolumn{2}{c||}{30fps $\rightarrow$ 60fps}    & \multicolumn{4}{c}{}                                           \\ \midrule
Metric   & PLCC          & SRCC         & PLCC          & SRCC         & PLCC          & SRCC         & PLCC          & SRCC         & PLCC          & KRCC          & SRCC         & RMSE           \\ \midrule
PSNR\textsubscript{EPE}                   & \underline{0.54}                  & 0.48                       & \underline{0.51}            & \underline{0.61}                        & 0.40                        & 0.37                       & 0.62             & 0.69                                   & 0.45     & 0.34      & 0.49      & 16.38     \\
SSIM\textsubscript{EPE}                   & \textbf{0.56}               & \textbf{0.51}              & \underline{0.51}            & \textbf{0.63}                     & 0.35                        & 0.37                       & 0.63             & 0.69                                   & 0.46     & 0.35      & 0.50      & 16.28     \\
FloLPIPS\textsubscript{EPE}               & 0.08                        & 0.04                       & 0.04                  & 0.12                              & 0.40                        & 0.31                       & 0.38             & 0.28                                   & 0.31     & -0.02     & -0.04     & 17.46     \\ \midrule
PSNR\textsubscript{TS}                & 0.45                        & 0.36                       & \underline{0.51}            & 0.50                              & 0.31                        & 0.30                       & 0.56             & 0.63                                   & 0.44     & 0.27      & 0.40      & 16.51     \\
SSIM\textsubscript{TS}                & 0.41                        & \underline{0.37}                 & 0.49                  & 0.60                              & {\ul 0.52}                  & \textbf{0.54}              & 0.63             & \underline{0.70}                                   & \underline{0.51}     & \textbf{0.38}      & \underline{0.54}      & \underline{15.83}     \\
FloLPIPS\textsubscript{TS}            & 0.04                        & 0.02                       & 0.15                  & 0.16                              & 0.35                        & 0.11                       & 0.19             & 0.06                                   & 0.14     & 0.05      & 0.07      & 18.17     \\ \midrule
PSNR\textsubscript{DIV}                   & 0.42                        & 0.33                       & 0.42                  & 0.50                              & 0.50                        & 0.50                       & \underline{0.69}             & 0.68                                   & 0.49     & {\ul{0.36}}      & 0.51      & 15.96     \\
SSIM\textsubscript{DIV}                   & 0.41                        & \underline{0.37}                 & 0.49                  & 0.60                              & {\ul 0.52}                  & \textbf{0.54}              & 0.63             & {\ul 0.70}                                   & {\ul{0.51}}     & \textbf{0.38}      & {\ul{0.54}}      & {\ul 15.83}     \\
FloLPIPS\textsubscript{DIV}               & 0.42                        & 0.32                       & \textbf{0.53}         & 0.55                              & \textbf{0.55}               & \underline{0.51}                 & \textbf{0.75}             & \textbf{0.73}                                   & \textbf{0.55}     & \textbf{0.38}      & \textbf{0.55}      & \textbf{15.32}            \\ \bottomrule
\end{tabular}
}
\end{table}

In Table \ref{tab:augmented-image} we show correlation results for the motion augmented image metrics. Comparing these results we can see that for the most part combining image information with motion derived measures increases performance over solely motion based metrics. However, using motion to weight FloLPIPS results in at times severe degradation in performance. It is likely more sensible moving forward to use the field of the derived motion metric to weight image measures spatially.

\subsection{Comparison of Computational Time for Metrics}

We now evaluate the computational load of our metrics compared to FloLPIPS. Table \ref{tab:run-time} shows run times in milliseconds for calculating each metric for a single 1080p frame. All metrics were evaluated on a 12th Gen Intel Core i7-12700K with an Nvidia RTX A4000. The highest computational load is incurred by VM-EPE due to the highly intensive process of the vector median filtering. Both EPE and DIV have lower runtimes than FloLPIPS with DIV being almost $3\times$ faster than FloLPIPS.

\begin{table}[]
\centering
\caption{\textit{Time required to calculate motion field and metric for a single 1080p frame in milliseconds.}}
\label{tab:run-time}
\resizebox{0.4\columnwidth}{!}{
\begin{tabular}{llll}
\toprule
         & \begin{tabular}[c]{@{}l@{}}Motion Est. \\ Time (ms)\end{tabular} & \begin{tabular}[c]{@{}l@{}}Calc. \\ Time (ms)\end{tabular} & \begin{tabular}[c]{@{}l@{}}Total \\ Time (ms)\end{tabular} \\ \midrule
FloLPIPS & 301.4                                                            & 3.7                                                        & 305.1                                                      \\ \midrule
EPE      & 224.6                                                            & 3.5                                                      & 228.1                                                      \\
TS   & 112.3                                                            & 860.2                                                    & 972.5                                                      \\
DIV      & 112.3                                                            & 0.1                                                      & 112.4                                                      \\
VM-EPE   & 112.3                                                            & 2788.3                                                      & 2900.6                                                     \\ \bottomrule
\end{tabular}}
\end{table}

The computational analysis reveals significant performance differences across the evaluated metrics. DIV emerges as the most computationally efficient metric, requiring only 112.4ms per frame—approximately 2.7$\times$ faster than FloLPIPS (305.1ms). This efficiency stems from DIV's simple divergence calculation, which requires only basic spatial derivatives of the motion field without complex filtering operations.

VM-EPE exhibits prohibitive computational costs at 2900.6ms per frame, making it nearly 10× slower than FloLPIPS. The vector median filtering process, which requires computing distance energies across spatial neighbourhoods for every motion vector, scales poorly with resolution and represents a significant computational bottleneck.

\begin{figure}
    \centering
    \includegraphics[width=0.45\linewidth]{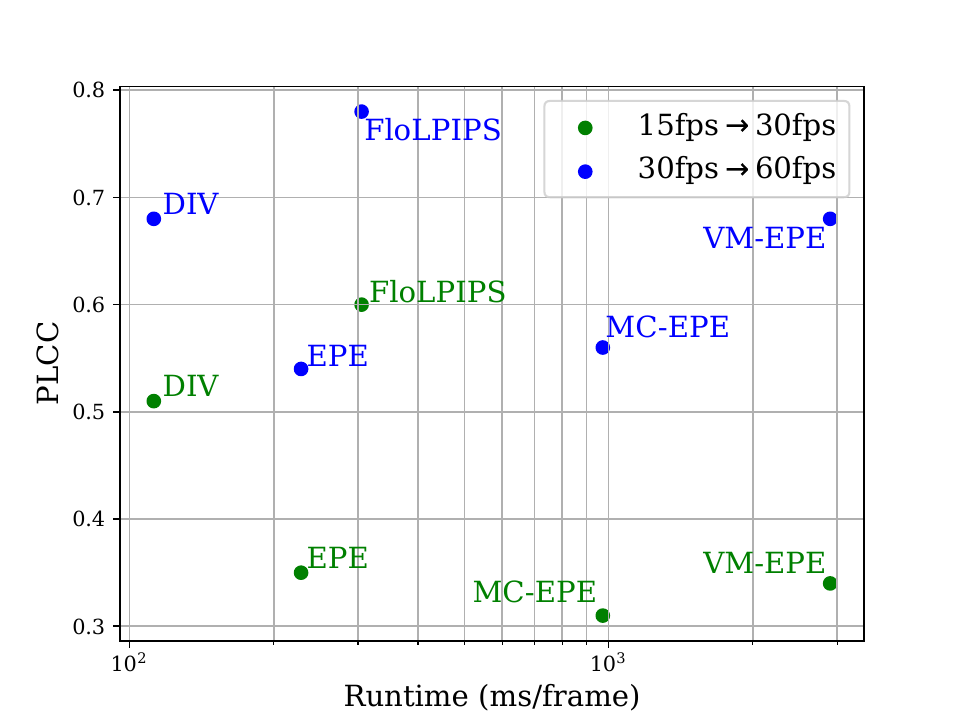}
    \caption{\textit{Pearson linear correlation coefficient (PLCC) for each metric plotted against metric runtime in milliseconds per frame (ms/frame) for a 1080p sequence.}}
    \label{fig:corr-time-tradeoff}
\end{figure}

Figure \ref{fig:corr-time-tradeoff} illustrates the performance-efficiency trade-off across both frame rate conversion scenarios. The analysis reveals that DIV offers good computational efficiency with only modest correlation degradation—achieving approximately 87\% of FloLPIPS's correlation performance while requiring only 37\% of the computational time. This represents an attractive trade-off for real-time applications where computational resources are constrained.

\subsection{Evaluating VFI Algorithms with Motion Metrics}
\label{subsec:eval-vfi}

In this section we use the motion-metrics we have developed to try and evaluate frame interpolation algorithms for motion consistency.

\begin{table}[]
\centering
\caption{\textit{Comparison of our motion metrics to commonly used metrics using BVI-VFI dataset interpolating from 30fps to 60fps at resolutions of 540p and 1080p. In each column, the algorithms scoring the best are \textbf{boldfaced} and the second best are \underline{underlined}.}\label{tab:eval_tab}}
\hfill
\resizebox{\columnwidth}{!}{%
\begin{tabular}{@{}ccccccccccc@{}}
\toprule
              & PSNR$\uparrow$           & SSIM$\uparrow$           & FloLPIPS$\downarrow$       & EPE$\downarrow$            & VM-EPE$\downarrow$          & TS$\downarrow$                & DIV$\downarrow$              & S$_{\text{dis}}$-S$_{\text{ref}}\downarrow$  & PSNR$_{\text{DIV}}\uparrow$ & SSIM$_{\text{DIV}}\uparrow$ \\ \midrule
ABME         & 37.5          & 0.906          & 0.085          & 2.59          & 1.23          & 0.028          & 0.023          & 0.022                                    & 36.5                   & 0.884                   \\
ACKMRF       & 32.2          & 0.672          & 0.254          & 3.06          & \underline{0.82}    & \underline{0.021}    & \underline{0.017}    & 0.025                                    & 31.6                   & 0.660                   \\
Average      & 35.0          & 0.794          & 0.077          & 4.81          & 1.74          & 0.025          & 0.021          & 0.042                                    & 34.3                   & 0.779                   \\
DVF          & 34.1          & 0.714          & 0.144          & 16.7         & 17.0         & 0.223          & 0.180          & 0.156                                    & 30.3                   & 0.654                   \\
FILM         & 38.3          & 0.907          & \textbf{0.044} & \underline{2.22}    & 1.08          & 0.032          & 0.026          & \underline{0.019}                              & 37.2                   & 0.882                   \\
LDMVFI       & 37.1          & 0.883          & 0.055          & 2.53          & \textbf{0.76} & \textbf{0.019} & \textbf{0.015} & 0.029                                    & 36.4                   & 0.868                   \\
Minterpolate & 36.7          & 0.875          & 0.069          & 3.23          & 1.79          & 0.037          & 0.030          & 0.028                                    & 35.6                   & 0.849                   \\
QVI          & 36.2          & 0.892          & \underline{0.052}    & 2.84          & 2.26          & 0.031          & 0.025          & \textbf{0.018}                           & 35.2                   & 0.869                   \\
re-sepconv   & 37.2          & 0.887          & 0.086          & 3.32          & 2.24          & 0.034          & 0.027          & 0.028                                    & 36.1                   & 0.862                   \\
Repeat       & 34.0          & 0.764          & 0.072          & 9.50          & 17.4         & 0.080          & 0.064          & 0.034                                    & 32.3                   & 0.731                   \\
RIFE         & 37.6          & 0.895          & 0.082          & 2.73          & 1.30          & 0.033          & 0.026          & 0.020                                    & 36.6                   & 0.870                   \\
softsplat    & 37.7          & 0.893          & 0.052          & 3.07          & 2.95          & 0.034          & 0.027          & 0.024                                    & 36.6                   & 0.868                   \\
STMFNet      & \textbf{38.8} & \textbf{0.919} & 0.070          & \textbf{2.06} & 1.19          & 0.026          & 0.021          & 0.019                                    & \textbf{37.9}          & \textbf{0.897}          \\
UPR-Net      & \underline{38.4}    & \underline{0.916}    & 0.065          & 2.42          & 1.03          & 0.027          & 0.021          & 0.021                                    & \underline{37.5}             & \underline{0.894}            \\
\bottomrule
\end{tabular}%
}
\end{table}

The evaluation in Table \ref{tab:eval_tab} reveals interesting disconnects between traditional image quality metrics and motion consistency measures across different algorithmic approaches. STMFNet and UPR-Net achieve the highest PSNR/SSIM scores but show relatively modest performance in motion-based metrics, suggesting these algorithms prioritise pixel-level reconstruction accuracy over temporal flow consistency.

A notable pattern emerges with LDMVFI (a diffusion-based model), which ranks first across multiple motion metrics (VM-EPE=0.76, TS=0.019, DIV=0.015) despite achieving only moderate PSNR/SSIM scores. This disconnect suggests that diffusion-based approaches may inherently produce more temporally coherent motion fields, possibly due to their generative nature and implicit regularization during the denoising process, even when this doesn't translate to improved pixel-wise similarity measures.

Similarly, ACKMRF, a Bayesian approach that explicitly enforces motion smoothness through Markov Random Fields, achieves second-best performance in multiple motion metrics (VM-EPE=0.82, TS=0.021, DIV=0.017) while recording the lowest PSNR/SSIM scores among learning-based methods. This pattern reinforces that algorithms with explicit temporal or motion regularization may sacrifice pixel-level fidelity to maintain motion consistency, highlighting a fundamental tension between reconstruction accuracy and temporal coherence.

FILM shows an interesting balance, achieving the best FloLPIPS score while maintaining competitive motion metrics, suggesting that perceptually-motivated training objectives may align better with temporal consistency than pure reconstruction losses. Conversely, DVF demonstrates a clear failure mode, maintaining reasonable PSNR but exhibiting catastrophically poor motion consistency (EPE=16.66), indicating severe temporal artifacts that pixel-level metrics fail to detect.

The results highlight fundamental algorithmic trade-offs that are invisible to traditional metrics. Algorithms optimizing for reconstruction fidelity (high PSNR/SSIM) may inadvertently introduce motion inconsistencies, while approaches with explicit temporal modeling (diffusion-based LDMVFI, Bayesian ACKMRF) appear to naturally preserve temporal coherence despite imperfect pixel reconstruction. This suggests that training objectives focused solely on pixel-level losses may be insufficient for producing temporally consistent interpolation.

\begin{figure}[]
    \centering
    \includegraphics[width=\linewidth]{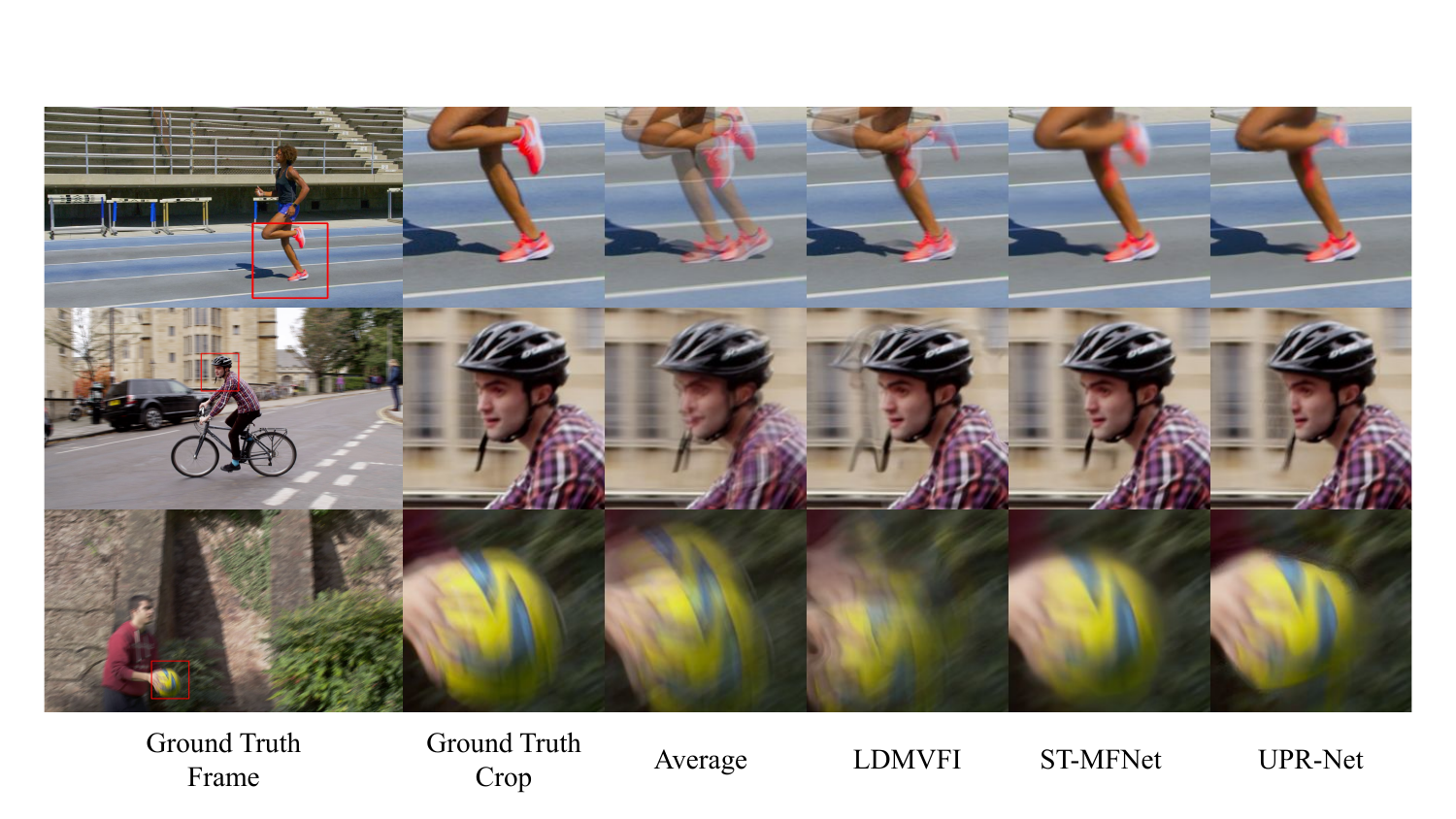}
    \caption{\textit{Representative frames taken from sequences in the BVI-VFI dataset and their interpolated version for 4 different algorithms. UPR-Net and ST-MFNet score well in terms of conventional metrics whereas LDMVFI scores well in terms of our motion-based metrics.}}
    \label{fig:crops}
\end{figure}

Figure \ref{fig:crops} provides visual context for these metric patterns. While the crops show that ST-MFNet and UPR-Net produce sharp, detailed interpolations that would score well in pixel-based metrics, the motion analysis reveals potential temporal artifacts that are not immediately apparent in static frame comparisons. The diffusion-based LDMVFI may sacrifice some fine detail reproduction in favor of maintaining more natural motion characteristics.

\section{Conclusions}
\label{sec:concl}
We have demonstrated that simple motion-based metrics can reliably estimate the quality of frame rate upsampled video sequences, offering computationally efficient alternatives to complex deep learning approaches. Our evaluation on the BVI-VFI dataset reveals that End Point Error (EPE) and Motion Field Divergence (DIV) achieve competitive correlation with human perceptual judgments while providing substantial computational savings—DIV processes frames 2.7$\times$ faster than FloLPIPS while maintaining 87\% of its correlation performance. Additionally, our motion-weighted image quality metrics, particularly SSIM\textsubscript{TS} and PSNR\textsubscript{DIV}, demonstrate meaningful improvements over baseline image metrics (PLCC increases by $\sim0.08$), showing that simple motion weighting can enhance traditional quality assessment approaches.

While our results show promise, several limitations deserve mention. The accuracy of our metrics depends on the reliability of the optical flow estimation. Errors in the estimated flow, especially in regions of occlusion or extreme motion, can propagate to the final metric values. Although motion-based metrics excel at capturing temporal inconsistencies, they may not always reflect subtle visual artifacts such as noise, colour shifts, or small texture details. Hence, using motion-based measures to weight image metrics spatially could yield better results. We plan to investigate these ideas in future work.

\acknowledgments 
 
This publication has emanated from research jointly funded by Taighde Éireann – Research Ireland, and Overcast HQ under Grant number EPSPG/2023/1515.

\bibliography{report} 
\bibliographystyle{spiebib} 

\end{document}